\documentclass[acmsmall]{acmart} 

\usepackage{float} 

\newcommand{\revision}[1]{{\textcolor{black}{#1}}}

\setcopyright{rightsretained}
\acmJournal{PACMHCI}
\acmYear{2025} \acmVolume{9} \acmNumber{2} \acmArticle{CSCW187}
\acmMonth{4} \acmDOI{10.1145/3711085}

\begin{document}

\title[The DSA Transparency Database]{The DSA Transparency Database: Auditing Self-reported Moderation Actions by Social Media}
\titlenote{\textcolor{red}{Article published in \textit{The 28th ACM Conference on Computer-Supported Cooperative Work and Social Computing -- CSCW'25}. DOI: \href{http://doi.org/10.1145/3711085}{10.1145/3711085}. Please, cite the published version.}}

\author{Amaury Trujillo}
\orcid{0000-0001-6227-0944}
\affiliation{\institution{Institute of Informatics and Telematics, National Research Council (IIT-CNR)}
\city{Pisa}
\country{Italy}
}
\email{amaury.trujillo@iit.cnr.it}

\author{Tiziano Fagni}
\orcid{0000-0003-1921-7456}
\affiliation{\institution{Institute of Informatics and Telematics, National Research Council (IIT-CNR)}
\city{Pisa}
\country{Italy}
}
\email{tiziano.fagni@iit.cnr.it}

\author{Stefano Cresci}
\orcid{0000-0003-0170-2445}
\email{stefano.cresci@iit.cnr.it}
\affiliation{\institution{Institute of Informatics and Telematics, National Research Council (IIT-CNR)}
\city{Pisa}
\country{Italy}
}

\renewcommand{\shortauthors}{Amaury Trujillo, Tiziano Fagni, and Stefano Cresci}

\begin{abstract}
  Since September 2023, the Digital Services Act (DSA) obliges large online platforms to submit detailed data on each moderation action they take within the European Union (EU) to the \textit{DSA Transparency Database}. From its inception, this centralized database has sparked scholarly interest as an unprecedented and potentially unique trove of data on real-world online moderation. Here, we thoroughly analyze all 353.12M records submitted by the eight largest social media platforms in the EU during the first 100 days of the database. Specifically, we conduct a platform-wise comparative study of their: volume of moderation actions, grounds for decision, types of applied restrictions, types of moderated content, timeliness in undertaking and submitting moderation actions, and use of automation. Furthermore, we systematically cross-check the contents of the database with the platforms' own transparency reports. Our analyses reveal that (\textit{i}) the platforms adhered only in part to the philosophy and structure of the database, (\textit{ii}) the structure of the database is partially inadequate for the platforms' reporting needs, (\textit{iii}) the platforms exhibited substantial differences in their moderation actions, (\textit{iv}) a remarkable fraction of the database data is inconsistent, (\textit{v}) the platform X (formerly Twitter) presents the most inconsistencies. Our findings have far-reaching implications for policymakers and scholars across diverse disciplines. They offer guidance for future regulations that cater to the reporting needs of online platforms in general, but also highlight opportunities to improve and refine the database itself.
\end{abstract}

\ccsdesc[500]{Human-centered computing~Empirical studies in collaborative and social computing}
\ccsdesc[500]{Information systems~Social networking sites}
\ccsdesc[500]{Social and professional topics~Governmental regulations}

\keywords{social media, content moderation, digital services act, online regulation}

\maketitle

\section{Introduction}
\label{sec:introduction}
Content moderation refers to the strategies and processes through which an online platform ensures that the content and behaviors it hosts adhere to the platform's self-defined standards and guidelines, and to legal regulations~\cite{gillespie2018custodians}.
Moderation interventions are the specific actions through which platforms enforce such strategies and processes.
Currently, a wide array of interventions are adopted, which involve removing content and users~\cite{jhaver2021evaluating,tessa2024beyond}, demoting, restricting, or otherwise reducing their visibility~\cite{le2021setting,trujillo2022make}, showing users warning messages~\cite{katsaros2022reconsidering} or attaching warning labels to problematic content~\cite{zannettou2021won}, and even acting on monetary rewards~\cite{ma2022m}.

However, despite their importance in maintaining safe and healthy online environments, the application of these interventions often sparks concerns. The controversies are multifaceted, such as considering the potential misuse of removals and reduced visibility decisions as a form of censorship that threatens users' freedom of speech~\cite{myers2018censored}. Furthermore, transparency in content moderation is often questioned, as the processes behind moderation decisions ---whether executed by black-box AI algorithms or human moderators--- are sometimes perceived as opaque, or lacking clarity and consistency~\cite{gillespie2020content}. Issues of explainability, fairness, and the susceptibility of automated systems to biases further contribute to the ongoing debate~\cite{nogara2023toxic}, highlighting the delicate balance that platforms must strike between upholding community standards and legal obligations, and respecting user rights in the digital sphere~\cite{grimmelmann2015virtues}.

\subsection{The Digital Services Act}

In light of the predominant role held by large online platforms in controlling the flow of information and shaping user opinions, the European Union (EU) recently enacted a strategic response to address the intricate challenges of regulating online spaces. This response also meets the growing demand for legislative intervention to reform the existing model centered on platform self-regulation, wherein social media have long been largely autonomous in defining and implementing their own rules and procedures. The Digital Services Act (DSA) is the regulation officially introduced by the EU on October 2022 to modernize its legal framework governing digital services~\cite{eu2020DSA}. Its aim is to create a safer and more accountable online environment for users within the EU, by addressing a range of issues related to online platforms, including content moderation, user safety, and the responsibilities of digital service providers~\cite{turillazzi2023digital, church2023digital}. The DSA specifically includes several articles with important implications for online content moderation, the most relevant of which are briefly presented in the following.
\begin{itemize}
    \item \textbf{Article 33} designates those platforms with more than 45M users in the EU as either \textit{very large online platforms} (VLOPs) or \textit{very large online search engines} (VLOSEs). Due to their strategic role, these must comply with the most stringent rules of the DSA. As of July 2024, the EU has designated 22 VLOPs and 2 VLOSEs.\footnote{\url{https://digital-strategy.ec.europa.eu/en/policies/list-designated-vlops-and-vloses}} \item \textbf{Article 15} demands that VLOPs and VLOSEs release periodic \textit{transparency reports}. These must contain updated aggregated figures on active recipients,\footnote{Those users who either requested hosting of information or were exposed to information hosted on the platform.} content removals, user appeals, use of automation and AI in moderation, and timeliness of interventions, among other information. \item \textbf{Article 17} obliges VLOPs and VLOSEs to submit a clear and specific \textit{statement of reasons} (SoR) for each of their moderation interventions. Each SoR must provide detailed and timely information on the intervention, its legal ground, and the content to which it was applied. Since February 2024 this article applies to all online platforms independently of their size. \item \textbf{Article 24(5)} establishes that the European Commission must set up and maintain a database of SoRs. On September 2023 the EU established the \textit{DSA Transparency Database}: an open, standardized, and centralized repository of all SoRs submitted within the scope of the DSA.
\end{itemize}

Collectively, these and all other DSA articles intend to set the stage for a transparent and harmonized approach to content moderation, underscoring the responsibilities of online platforms while ensuring consistency and fairness. The establishment of the DSA Transparency Database (henceforth \texttt{DSA-TDB}), in particular, marks an important milestone toward enforcing accountability, as it represents an unprecedented resource of self-reported data that enables for the first time the possibility to track, scrutinize, and compare real-world platform moderation actions.

\begin{figure}[t]
    \centering
    \includegraphics[width=1\textwidth]{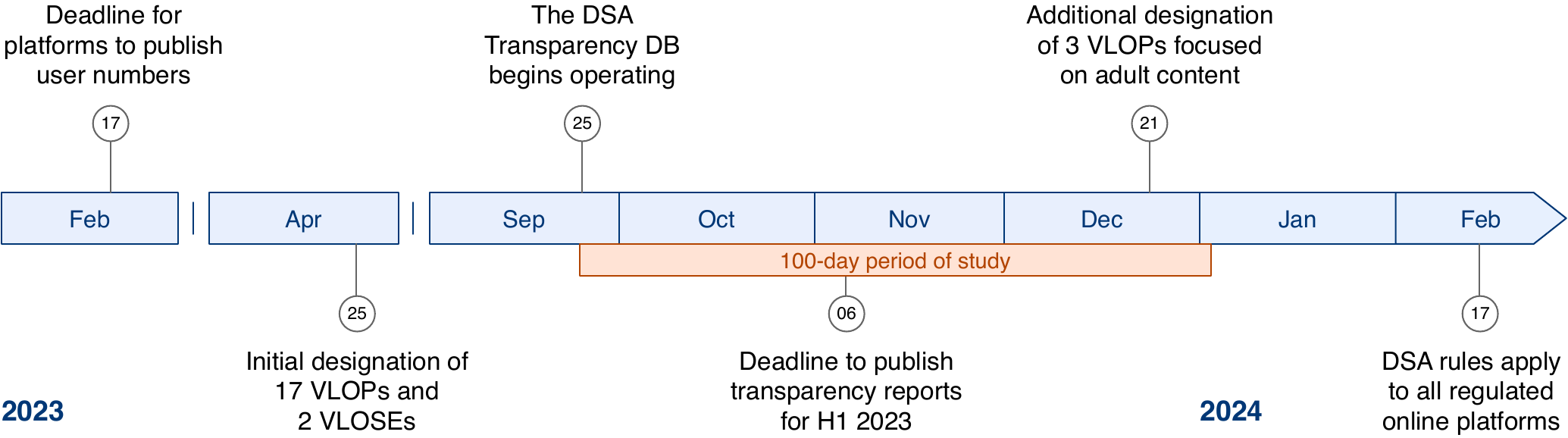}
    \Description{17 February 2023 was the deadline for platforms to publish user numbers. On 25 April the European Commission designated 17 VLOPs and 2 VLOSEs. On 25 September 2023 the DSA Transparency Database begins operating. 5 November was the deadline to publish transparency reports for the first semester of 2023. On 21 December the European Commision designated another 3 VLOPs, all of which focus on adult content. From 17 February 2024 the DSA rules are applied to all regulated online platforms, not only those deemed very large.}
    \caption{Timeline of relevant events concerning the DSA Transparency Database.}
    \label{fig:dsa_timeline}
\end{figure}

\subsection{Research Questions and Contributions}

Herein, we carry out an in depth quantitative analysis of very large social media moderation practices, as reflected by their SoR submissions to the \texttt{DSA-TDB}. We analyze all 353.12M SoRs submitted by eight platforms during the first 100 days of the \texttt{DSA-TDB}, providing an early yet thorough look into (\textit{i}) the overall frequency of moderation interventions by very large social media, (\textit{ii}) their grounds for decision, (\textit{iii}) the types of applied restrictions, (\textit{iv}) the types of moderated content, (\textit{v}) the timeliness of their interventions, and (\textit{vi}) their use of automation in content moderation. Figure~\ref{fig:dsa_timeline} depicts this 100-day period of study (orange-colored) in the context of relevant events related to the \texttt{DSA-TDB}. In carrying out these analyses, we specifically focus on answering the following research questions.

\begin{description}
    \item[RQ1:] \textit{To what extent do social media provide adequate and detailed information in the SoRs they submit?} \end{description}
Article 17 of the DSA specifies that each submitted SoR should be clear, specific, detailed, and timely. RQ1 evaluates the extent to which the different platforms aligned with these criteria. Then, other than shedding light on the adequacy and informativeness of the submitted SoRs, the analysis also uncovered some inherent limitations in the structure of the \texttt{DSA-TDB} that hinder platforms from fully meeting the goals of the DSA. 

\begin{description}
    \item[RQ2:] \textit{How consistent and coherent are the SoRs submitted by each social media?}
\end{description}
Articles 15 and 24(5) of the DSA establish two transparency mechanisms: transparency reports and the \texttt{DSA-TDB}. By cross-checking information submitted by each platform to the \texttt{DSA-TDB} and by comparing such information with those contained in their DSA Transparency Reports, the analyses entailed by RQ2 reveal the degree to which each platform submitted consistent and coherent information. In particular, we found striking inconsistencies in the information submitted by some of the platforms.

\begin{description}    
    \item[RQ3:] \textit{How do social media differ in their self-reported moderation actions?} \end{description}
Assuming that the self-reported information contained in the \texttt{DSA-TDB} are reliable, their analysis sheds lights on the moderation practices of the different platforms. Our results to RQ3 reveal common patterns and marked differences between platforms, along multiple fundamental dimensions of content moderation, including the timeliness of interventions and the use of automation.

\subsubsection{Significance}
To the best of our knowledge, this is the first in-depth, platform-wise comparative analysis of moderation actions over an extended time period, in a vast social media ecosystem, at a continental scale. As such, the implications of our work are many and various. On the one hand, our results allow drawing insights into the moderation practices of different platforms, for example, by identifying common patterns and striking differences. We also uncover incoherent data submitted by some platforms, which raises concerns on the reliability of some of the information contained in the \texttt{DSA-TDB}. On the other hand, our analysis also sheds light on some of the challenges implied by imposing a unified reporting structure to a large number of diverse platforms. Our findings can thus inform future developments of the \texttt{DSA-TDB}, so as to possibly overcome the limitations of the current structure.

 \section{Related Work}
\label{sec:related-work}
We first present an overview of the \emph{implications of the DSA} for online regulation, particularly on content moderation. We then consider the critical aspect of \emph{platform transparency} for content moderation, both in the form of reports by platforms themselves and in the form of access to platforms' data for independent analysis. Finally, we discuss a couple of \emph{precursory analyses} of the \texttt{DSA-TDB} in its earliest stage, and how the present work improves upon these.

\subsection{The implications and Impact of the DSA}
The DSA marks a crucial advancement in EU digital services regulation, aiming to enhance user protection in an ever-increasing digital world. It focuses on improving online safety, transparency, and provider accountability \cite{wilman2022digital, cauffman2021new}.
Moreover, the DSA contains stringent measures ---particularly for VLOPs and VLOSEs--- to reduce risks and foster safer platforms for millions of users that could inspire other jurisdictions to update their current norms~\cite{husovec2024rising}.
For these reasons, the DSA has sparked scholarly debate on its potential implications and impact, in Europe and beyond.

For instance, ~\citet{nunziato2023digital}, along with~\citet{church2023digital}, explored the global impact that the DSA could have on social media content moderation. They suggest that the stringent requirements of the DSA could set global Internet moderation norms due to the high cost of implementing different systems for different jurisdictions. Given the attractiveness and power of the European single market, platforms are likely to use worldwide their already DSA-compliant systems. This is also likely to create some friction with other important markets, such as the United States of America, as the more stringent approach by the EU contrasts the approach by the USA, which often prioritizes freedom of speech.
In particular, the pivotal Section 230, enacted in 1996 as part of the USA Communications Decency Act, states that online platforms are not to be treated as publishers, and are thus not liable for content created by users within the platform~\cite{section230}. Such provision has been both heralded as a saviour of free speech and criticized as a shield for scoundrels~\cite{ardia2009free}.
The DSA also safeguards service providers in a similar manner in Articles 14 and 15. However, it is much more stringent on the obligations of service providers and in the kinds of contents deemed illegal and the actions required.
In this view, \citet{schneider2023effectiveness} assessed the potential for harm reduction caused by the DSA by means of self-exciting point processes. Their analysis, based on real-world data, indicates that the DSA has the potential to standardize and effectively regulate content moderation, with an emphasis on rapid action against highly harmful content.

Instead of focusing on the broad implications of the DSA as a whole, other scholars investigated its possible effects in more specific contexts. For example, \citet{leerssen2023end} studied the lawful basis for inherently opaque moderation interventions such as demotions and ``shadow banning''~\cite{le2021setting}, revealing a strong tension between platforms' moderation needs and the transparency-enhancing framework of the DSA. Finally, \citet{strowel2023digital} discussed the implications of the DSA in combating online disinformation, underscoring platform transparency requirements and their potential in curbing misleading content. The work also reflected on the challenges and effectiveness of these obligations, proposing enhancements to the regulatory framework of the DSA regarding enforcement and compliance.

\subsection{Platform Transparency on Content Moderation}

Social media platforms, crucial as they are in shaping public discourse, face scrutiny over their content moderation practices. Balancing freedom of expression with digital civil discourse is challenging \cite{denardis2015internet, gillespie2018custodians}, and criticism arises due to a perceived lack of transparency and biases in moderation \cite{jhaver2019did, myers2018censored}. Far-right users claim censorship \cite{jahanbakhsh2021exploring, myers2018censored}, while left-leaning users point to policy enforcement inconsistencies \cite{shen2019discourse}. In addition, marginalized communities often face disproportionate moderation \cite{haimson2021disproportionate,thach2022visible}. The evolution from manual to automated moderation \cite{zhang2021automatic, annamoradnejad2023automatic} further complicates matters, with mixed user opinions on their fairness and bias \cite{lyons2022s, gonccalves2023common, nogara2023toxic}. Therefore, transparent content moderation practices by online platforms are of paramount importance. In this regard, the DSA provides three important mechanisms: periodic transparency reports, granting researchers access to platform data, and the new centralized transparency database ---the subject of the present study.

Transparency reports are an accountability instrument to detail user and authority notifications, content removals, and accuracy of a platform moderation system. These reports ---already mandatory for VLOPs and VLOSEs under the DSA--- provide critical insights into digital governance, enhancing public understanding and regulatory oversight~\cite{buri2021digital,eu_transparency_reports}.
For example, \citet{urman2023transparent} carried out a detailed analysis of the transparency reports released by large online platforms pursuant to Article 15 of the DSA. They found partial compliance with the Santa Clara Principles 2.0~\cite{santaclaraprinciples2020santa} on transparency and accountability in content moderation, with Snapchat being fully compliant and Google excelling in state-involved reporting. Their findings also highlighted widespread non-compliance in flagging processes, underscoring a need for enhanced transparency within the DSA.

Having access to data posted on social media through specific APIs or historical repositories is also important toward moderation transparency. Arguably, the then free Twitter Researcher API~\cite{chen2022twitter} and Reddit Pushshift archive~\cite{baumgartner2020pushshift} were among the most open and widely used data for social media analysis. For several years, these enabled researchers to investigate the effects of moderation policies of their respective platforms.
For instance, several studies have investigated the impact of major moderation interventions on user behavior in Reddit, at the community~\cite{trujillo2022make} and individual levels~\cite{trujillo2023one}, inside the platform as well as across platforms~\cite{horta2021platform}. Interestingly, in several cases ---and particularly in the case of political communities--- such moderation interventions had mostly mixed effects~\cite{chandrasekharan2022quarantined, shen2022tale}.

Moderation interventions by Twitter were also found to have mixed effects. For instance, deplatforming controversial influencers was followed by a general decrease in conversations and toxicity, but also by an increase in offensive content among certain users~\cite{jhaver2021evaluating}. In the case of Trump's ban after the Capitol attack, there was even increased polarization and ideological differences~\cite{alizadeh2022content}.
More recent studies have shown that Twitter was also the target of content manipulation in the context of the Ukraine invasion and the French presidential election in 2022~\cite{pierri2023does}.
Similar phenomena has been observed in historically less open platforms, such as Facebook and Instagram, across different topics, ranging from political issues~\cite{kalsnes2021hiding, goldstein2023understanding} to groups focused on unhealthy behaviours~\cite{chancellor2016thyghgapp, gerrard2018beyond}.

In early 2023, due to changes in management and in part to the unauthorized use of social media content to train generative AI models, the Twitter API ceased to be free\footnote{\url{https://twitter.com/XDevelopers/status/1621026986784337922}} (even for research) and Reddit severely restricted access to its API.\footnote{\url{https://redd.it/12qwagm/}} Indeed, both platforms, as well as others, changed their terms of service with respect to data access and collection.
Consequently, many social media analysis projects were halted, significantly hindering the transparency and accountability of social media. \revision{In this respect, Article 40 of the DSA establishes a provision for granting vetted researchers access to VLOPs and VLOSEs data, provided that certain conditions are met~\cite{eu2020DSA}. The overarching goal is that of understanding how society is shaped by these platforms, as well as supporting their oversight of DSA compliance. However, some of the very large platforms have just recently set up the mechanisms to access EU content, while independent investigations documented widespread hurdles at accessing platform data under Article 40 of the DSA, raising serious concerns on platform compliance with the regulation~\cite{jaursch2024dsa}. Therefore, the effect of such provision on online moderation research remains to be seen.}

\subsection{Precursory Analyses of the DSA Transparency Database}

The \texttt{DSA-TDB}, an unprecedented potential data trove regarding moderation actions, has piqued much scholarly interest since its announcement. \revision{Even at this early stage, a few works have provided preliminary analyses of the initial data submitted by large social media platforms to the database. \citet{kaushal2024automated} performed a multidisciplinary legal and computational analysis of a data sample of 131M SoRs from all VLOPs in the database, including shopping platforms, social media, app stores, and service platforms. They concluded that while the database provides some insight into platform moderation actions, its reliance on self-reported data limits transparency and legal certainty, revealing inconsistencies, potential manipulation, and insufficient compliance with the DSA's objectives, necessitating complementary transparency mechanisms and robust enforcement by the European Commission.
A different study investigated the degree of variability of content moderation practices for the main social media platforms across the EU member states, with a particular focus on the use of automated moderation~\cite{papaevangelou2024content}. By analyzing four months of data, the study found inconsistent information and a great deal of heterogeneity between the different platforms, and even for certain platforms across different EU member states.} The study by \citet{dergacheva2023one} only took into consideration submissions made during the very first day of the database. However, in spite of the extremely limited breadth of the analysis, they found much heterogeneity in the reported degrees of automation, applied visibility measures, and types of content moderated by various online platforms on a single day.

In a work more similar to our present contribution, \citet{drolsbach2023content} examined the first 60 days of data from the database, revealing relevant disparities in moderation practices across different platforms. Their findings underscored the predominance of automated moderation, with X (formerly Twitter) being the main exception, and highlighted some inconsistencies in implementing the DSA guidelines. Differently from our study, the work by \citet{drolsbach2023content} is mostly descriptive and it does not systematically take into consideration the differences among platforms. For instance, a  regression analysis was conducted regarding the use of automated means for moderation, which however conflated different levels of automation and data from various platforms. Given that TikTok represented more than 60\% of the data, their resulting model is heavily influenced by the self-reported practices of a single platform, with very different dynamics compared to the other seven social media.

On the other hand, our research presents a different approach to the database and includes several aspects not covered in any of the previous analyses. Firstly, we investigate the initial 100 days of the database, providing a much more comprehensive and robust time frame. Secondly, we conduct a thorough platform-wise analysis of moderation actions, taking into consideration the potential distinctiveness of each social media platform. Thirdly, our study provides a deeper assessment of the data submitted to the database, specifically addressing questions about the adequacy, consistency, and timeliness of submitted SoRs. Lastly, we also carry out a systematic coherence cross-check of the database against each platforms' public DSA Transparency Reports, which allows to identify discrepancies and inaccuracies in the reported data. Our approach not only uncovers worrisome inconsistencies, but also critically evaluates the effectiveness and limitations of the current database schema, while suggesting possible enhancements.
 \section{Overview of the DSA Transparency Database}
\label{sec:overview}

The main entity in the \texttt{DSA-TDB} is the \emph{statement of reasons} (SoR) ---specific information regarding the removal or restriction to a certain content by providers of hosting services, intended to empower users to understand and challenge content moderation decisions. Herein, we analyze all SoRs submitted during the initial 100 days since the establishment of the database, from 25 September 2023 to 02 January 2024. The data, documentation, and source code of the \texttt{DSA-TDB} are publicly available under permissive licenses on its official website.\footnote{\url{https://transparency.dsa.ec.europa.eu/}} For our analyses, we used the ``light'' version of the daily archives, released as compressed CSV files with 33 fields. With respect to the the full database schema, the ``light'' version excludes four lengthy fields: three free text attributes and one territorial scope attribute. Based on a brief analysis of these four attributes on a sample of a few days of the ``full'' archive, we decided for the ``light'' version, as we considered them uninformative in answering our research questions.

Out of the 17 VLOPs initially designated by the DSA,\footnote{\url{https://ec.europa.eu/commission/presscorner/detail/en/IP_23_2413}} 16 have submitted at least one SoR. Wikipedia is the only VLOP that did not make any submission to the \texttt{DSA-TDB} during the initial 100-day period. In this time, a total of 1.4861B SoRs were submitted. The largest submitter was by far Google Shopping, with more than a billion SoRs (67.8\% of the total), and the smallest was Zalando, with merely 9 SoRs submitted. Given our focus on social media ---platforms centered on user generated content and social networking--- we selected the following eight platforms of interest for this study: Facebook, Instagram, LinkedIn, Pinterest, Snapchat, TikTok, X (formerly Twitter), and YouTube. With 353.12M SoRs, these platforms account for 23.8\% of the whole database.

In addition to the \texttt{DSA-TDB}, we also took into account data available on the most recent DSA Transparency Reports released by the eight selected social media (see Table~\ref{tab:dsa_report_links} in appendix \ref{appx:dsa_report_links}). These compulsory reports mainly cover the first semester of the year (H1~2023) and were all made publicly available around October 2023 on the websites of the respective platforms. In particular, we manually extracted the average monthly authenticated active recipients (MAAR) of each platform for H1~2023. MAAR are registered and logged-in active recipients of the service. In the context of moderation, the number of MAAR is more relevant than that of all active recipients, because an active recipient that is not authenticated is usually not allowed to create or generate content on a platform, which is in turn the main target of moderation decisions. Moreover, in the absence of detailed data on the total content created for this period, the average MAAR is a useful, if faulty, metric to relativize the number of SoRs by platform. Table~\ref{tab:overview} presents an overview of these metrics for the eight selected social media.

\begin{table}[t]
\caption{Overview of very large social media platforms, ordered by the number of statements of reasons (SoRs) per average monthly authenticated active recipients (MAAR).}
\label{tab:overview}
\small
\begin{tabular}{lrrrr}
\toprule
\textbf{platform} & \#SoRs & \%SoRs & \#MAAR & $\frac{\text{\#SoRs}}{\text{\#MAAR}}$ \\ 
\midrule\addlinespace[2.5pt]
TikTok & $184.772$M & $52.33\%$ & $135.9$M & $1.360$ \\ 
Pinterest & $60.297$M & $17.08\%$ & $124.0$M & $0.486$ \\ 
Facebook & $79.050$M & $22.39\%$ & $259.0$M & $0.305$ \\ 
YouTube & $19.267$M & $5.46\%$ & $445.8$M & $0.043$ \\ 
Instagram & $8.111$M & $2.30\%$ & $259.0$M & $0.031$ \\ 
Snapchat & $1.119$M & $0.32\%$ & $102.0$M & $0.011$ \\ 
X & $0.466$M & $0.13\%$ & $61.3$M & $0.008$ \\ 
LinkedIn & $0.038$M & $0.01\%$ & $45.2$M & $0.001$ \\ 
\bottomrule
\end{tabular}
 \end{table}
 \section{Analyses and Results}
\label{sec:results}
Upon exploration of the database, we realized that platforms had varying approaches in how they filled in and mapped the attributes of their SoR submissions, especially in terms of mandatory and optional fields. For instance, only 3 out of 16 VLOPs indicate SoR \textit{content language} ---an optional attribute--- with none of the social media doing so. Therefore, in this section we present only the aspects that we considered more pertinent in answering our research questions, namely: \textit{grounds for the decision} on an infringement, \textit{types of restriction} decisions, \textit{type of moderated content}, \textit{timeliness} of intervention application and communication to the \texttt{DSA-TDB}, and the \textit{use of automated means} for moderation. When presenting and discussing these aspects, we make explicit reference to the corresponding section of the official documentation\footnote{\url{https://transparency.dsa.ec.europa.eu/page/documentation}} at the time of writing. This allows us to map each analysis to the corresponding field(s) in the dataset that we utilized, providing a clear reference to the documentation to enhance understanding of the field(s) semantics. Furthermore, by comparing our findings with the documentation, we gain valuable context on how the platforms have used the \texttt{DSA-TDB} fields, in relation to their intended or expected use. As a final validation step, we compare the data submitted by each social media to the \texttt{DSA-TDB} with the respective claims from their DSA Transparency Reports, so as to detect possible inconsistencies. 

\subsection{Grounds for Decision}
There are several attributes to indicate the legal or contractual grounds relied on in taking a moderation intervention decision, and to codify the infringement type. We first considered two grounds attributes (\S 11 Decision Grounds in the official documentation): a mandatory \emph{decision ground} (illegal, incompatible), and an optional attribute to indicate that a content is both incompatible with the terms of service and illegal. We then analyzed three attributes to codify infringements (\S 16 Category \& Specification): \textit{category} (mandatory high-level attribute from a predefined set of 14 categories), \textit{category addition} (optional list from the same set of categories), and \textit{category specification} (optional keyword list from a set of 55 predefined values).
Concerning the decisions grounds, for almost every platform \emph{incompatible content} was the overwhelming majority (more than 99\%), up to all instances in the case of LinkedIn. The exception was X, which indicated \emph{illegal content} for all of its SoRs.
Curiously, only Snapchat indicated an overlap of incompatible and illegal content in 12.2\% of their SoRs. None of the other social media platforms used this attribute.

\begin{figure}[t]
    \centering
    \includegraphics[width=.55\textwidth]{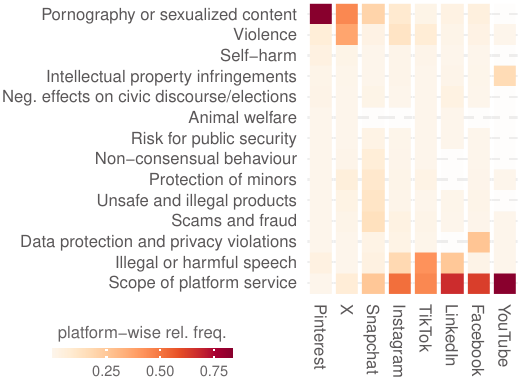}
    \Description{A heatmap shows that platforms can be grouped into two groups with similar distributions,  plus a single-platform group that does not share a similar distribution to others in terms of infringement categories. A first group is composed on Pinterest and X, in which sexual and violent content are the most common infringement categories. A second group is composed of YouTube, Facebook, LinkedIn, TikTok, and Instagram, for which the category scope of platform service was the most common. A third group is composed of Snapchat, which unlike the other platforms, does not present a dominant category.}
    \caption{Platform-wise distribution of infringement categories of statement of reasons (SoR), seriated using Manhattan distance across platforms and categories, so that similar platforms are positioned close to one another.}
    \label{fig:sor_categories_trellis}
\end{figure}

Figure~\ref{fig:sor_categories_trellis} provides a comparative analysis of the platforms based on the infringement categories of their SoRs. As shown, there exist three groups of similar platforms:
\begin{enumerate}
    \item those in which \emph{scope of platform service}\footnote{This category indicates content moderated because deemed outside of the scope of the platform service. It includes content with age-specific restrictions, with geographical or language requirements, disallowed goods and services, and nudity.} was the most common infringement (YouTube, Facebook, LinkedIn, TikTok, and Instagram);
    \item those in which \emph{pornography or sexualized content} was the most common infringement (Pinterest and X);
    \item a single-platform group (Snapchat), for which the previous categories are the most common, along with \emph{scams and fraud}.
\end{enumerate}

Curiously, \emph{category addition} was merely used by Pinterest and X, but in less than 0.04\% of their SoRs. 
As for keywords, only half of the social media platforms used them for \emph{category specification}: Pinterest, Snapchat, X, and YouTube. However, these are not always informative. For instance, circa 83\% of YouTube's SoRs have a keyword ---all of them \emph{other}--- which YouTube explained by specifying that ``[their] mapping has not yet been completed''. Only 11\% of Snapchat's SoRs have a keyword, 87\% of these being \emph{sexual abuse material} and the rest mainly \emph{online bullying intimidation}. About 88\% of Pinterest's SoRs have a keyword (with a few having two), predominantly \emph{adult sexual material}. Interestingly, all of X's SoRs have a keyword, with a few having two or three. In total X used 17 distinct keywords, more than any other social media. However, in almost half (43\%) of its SoRs the keyword is \emph{other} (detailed as a policy violation), while in 49\% it is \emph{adult sexual material}. In the remaining SoRs the keywords are primarily \emph{child sexual abuse material}, \emph{nudity}, and \emph{goods/services not permitted}.

\subsection{Types of Restriction Decisions}

\begin{table}[t]
    \caption{Platform-wise distribution of decision types, ordered alphabetically and by type mean across platforms.}
    \label{tab:decision_kind}
    \small
\setlength{\tabcolsep}{3pt}
\begin{tabular}{lrrrr}
\toprule
\multicolumn{1}{l}{} & \multicolumn{4}{c}{\textbf{decision type}} \\
\cmidrule(lr){2-5}
\multicolumn{1}{l}{} & \textit{visibility} & \textit{account} & \textit{provision} & \textit{monetary} \\
\midrule\addlinespace[2.5pt]
Facebook & $42.87\%$ & $57.13\%$ & -- & -- \\
Instagram & $44.11\%$ & $55.89\%$ & -- & -- \\
LinkedIn & $99.77\%$ & -- & $0.23\%$ & -- \\
Pinterest & $99.95\%$ & $>0\%$ & $0.04\%$ & -- \\
Snapchat & $80.15\%$ & $19.85\%$ & -- & -- \\
TikTok & $98.30\%$ & $1.15\%$ & $0.55\%$ & -- \\
X & $71.44\%$ & $6.58\%$ & $21.98\%$ & -- \\
YouTube & $98.04\%$ & -- & $1.96\%$ & -- \\
\midrule
\textbf{type mean} & $79.33\%$ & $17.58\%$ & $3.09\%$ & -- \\
\bottomrule
\end{tabular} \end{table}

\begin{figure}[t]
    \centering \includegraphics[width=.55\textwidth]{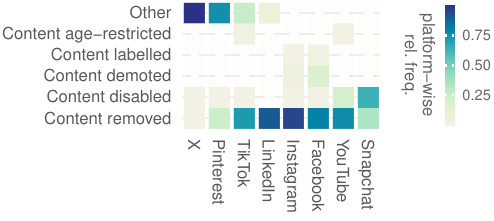}\Description{A heatmap shows that platforms can be grouped into two groups with similar distributions,  plus a single-platform group that does not share a similar distribution to others in terms of decisions on visibility. A first group is composed on Pinterest and X, in which other is the most common. A second group is composed of TikTok, LinkedIn, Instagram, Facebook, and YouTube, for which the visibility content removed was the most common. A third group is composed of Snapchat, which has content disabled as the most common visibility.}
    \caption{Platform-wise distribution for decisions of the \emph{visibility} type, seriated using Manhattan distance across platforms and sub-types, so that similar platforms are positioned close to one another.} \label{fig:sor_decision_visibility_subkinds_trellis}\end{figure}

There are four main types of restriction decisions (\S 3 The type of restriction(s) imposed): \emph{account} (suspension, termination), \emph{monetary} (suspension, termination, other), \emph{provision} (partial/total suspension/termination of the service), and \emph{visibility} (restriction for the content). Platforms must indicate a single sub-type of one of the four types, but these types are not mutually exclusive. In other words, a SoR can contain multiple types of restrictions, with at least one of them being mandatory. Nonetheless, all of the social media platforms only set a single restriction type. As detailed in Table~\ref{tab:decision_kind}, most were in \emph{visibility} (79.3\% across platforms), followed by \emph{account} (17.6\%) and \emph{provision} (3.1\%). Surprisingly, there was not a single SoR with a \emph{monetary} restriction decision. For \emph{visibility} sub-types, shown in Figure~\ref{fig:sor_decision_visibility_subkinds_trellis}, we noticed a similar platform grouping to that of infringement categories:
\begin{enumerate}
    \item those in which \emph{content removed} is the most common sub-type (Facebook, Instagram, LinkedIn, TikTok, and YouTube);
    \item those in which \emph{other} is the most common sub-type (X and Pinterest);
    \item Snapchat by itself, with \emph{content disabled} being its most common sub-type.
\end{enumerate}

\subsection{Content Type}
\begin{figure}[t]
    \centering
    \includegraphics[width=1\textwidth]{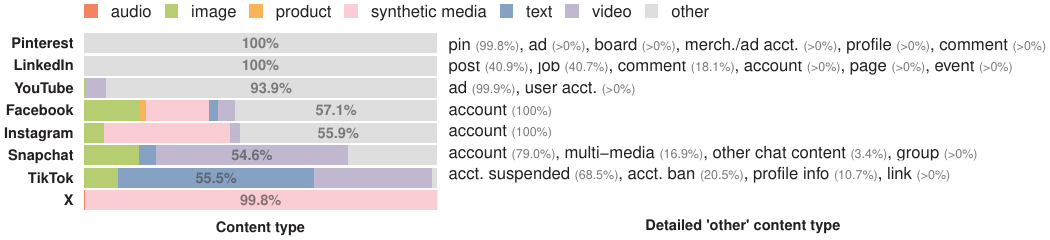}
    \Description{Stacked barchart ordered by distribution of moderated content type. Both Pinterest and LinkedIn have 100\% of their moderated content as other, followed by YouTube with 93.9\%. Both Facebook and Instagram have slightly more than half of their content type as other. On the other hand, most of Snapchat's moderated content type is video, whereas for TikTok was less than half, with text being more than half. X is the most dissimilar, with almost all of its moderated content being synthetic media.}
    \caption{Distribution of moderated content type by platform. For the detailed \emph{other} content type (right side of the figure), we simplified the original free text labels for space reasons.}
    \label{fig:sor_content_type_barchart}
\end{figure}

The required attribute \emph{content type} (\S 2 Specifications of the content affected by the decision) is a list with one or more of the following predefined values: \emph{app}, \emph{audio}, \emph{image}, \emph{product}, \emph{synthetic media} (e.g., content generated via AI), \emph{text}, \emph{video}, and \emph{other}. Despite the possibility to indicate multiple values, as also mentioned in the official documentation with multiple examples, almost all platforms only specified a single value for all of their SoRs. The only exception is YouTube, which however indicated multiple values only in a handful of cases ($<$0.01\%). We found that social media platforms greatly differ in how they specify the content type of their SoRs, as shown on the left side of Figure~\ref{fig:sor_content_type_barchart}. Pinterest and LinkedIn only set content type \emph{other}, and did not map any of their internal types to one or more of the predefined values in the \texttt{DSA-TDB}. YouTube also had a remarkable share of \emph{other} content (93.9\%), prevalently ads. Facebook and Instagram have similar distributions, with most content being \emph{other} (more than 55\%), all detailed as accounts, which is consistent with their distributions of decision types in Table~\ref{tab:decision_kind}. Incidentally, Facebook was the only social media with content type \emph{product} (1.8\%). More than half of Snapchat's moderated content was \emph{video}, with \emph{other} content (24\%) being mainly accounts. Still, 16.6\% of \emph{other} was labelled as multi-media, instead of using multiple predefined values as prescribed by the \texttt{DSA-TDB}. Interestingly, TikTok's moderated content was mostly \emph{text} (55.5\%) and not \emph{video} (33.5\%), with a small share of \emph{other} content (1.3\%), primarily accounts. Unexpectedly, moderated content by X was overwhelmingly \emph{synthetic media} (99.8\%), with the second being \emph{audio} (0.01\%) ---the only social media to use this type--- and only a handful of \emph{text}.

\subsection{Timeliness}
\begin{figure}[t]
    \centering
    \includegraphics[width=.625\textwidth]{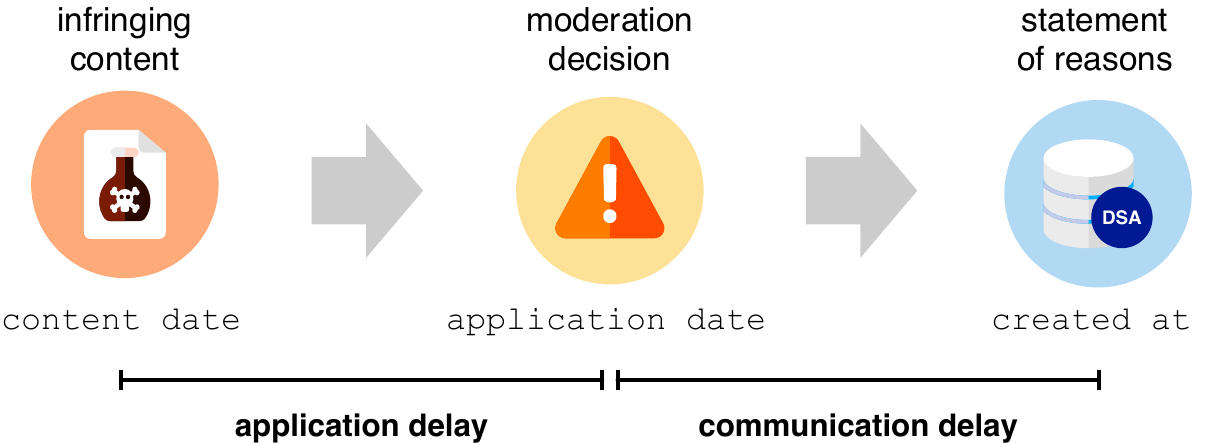}
    \Description{The diagram illustrates three events: creation of infringing content, application of moderated decision, and creation of statement of reasons. The application and communication delays are the respective lags between these three events.}
    \caption{We analyzed the timeliness of submitted moderation actions based  on: lag between the creation of the infringing content and the application of the moderation decision (\emph{application delay}); and lag between this application and the communication to the \texttt{DSA-TDB}, which implies the creation of a statement of reasons (\emph{communication delay}). Labels in monospaced font refer to the database fields used for computing the delays.}
    \label{fig:delays_diagram}
\end{figure}

Artice 17 of the DSA requires that platforms communicate their SoRs without undue delay to allow close to real-time updates. Pinterest, Snapchat, and TikTok started submitting SoRs on the very day that the \texttt{DSA-TDB} was made available ---that is, on 25 September 2023. Facebook, Instagram, YouTube, and LinkedIn the following day. X was last in line and started a week later, on 3 October 2023. As depicted in Figure~\ref{fig:delays_diagram}, we then analyzed the delay with which each social media communicated their SoRs, measured as the difference (in days) between the date of application of the moderation intervention and the date of creation of the corresponding SoR. In addition, a prompt response by platforms in moderating an infringing content is arguably desirable. Hence, we also measured the application delay (in days) of the moderation decision with respect to the creation of the infringing content. To compute the communication and application delays we use the content creation date (\S 2.3 Date on which the content was created on the online platform), the moderation intervention application date (\S 4.1 Application date), and the SoR creation date, which is among the available metadata. 

\begin{figure}[t]
    \centering
    \includegraphics[width=1\textwidth]{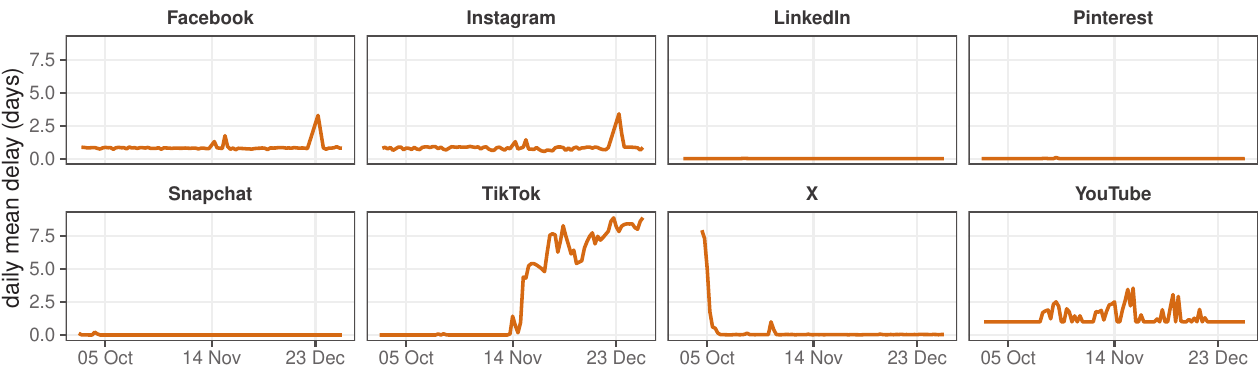}
    \Description{Faceted time series of daily mean delay by platform for the 100 days of the database. Facebook and Instagram have very similar time series, both presenting a relatively low delay but with a noticeable spike at the end. LinkedIn, Pinterest, and Snapchat have a minimal and steady delay. TikTok started with a minimal delay but halfway started to increase up to almost 8 days by the end. X started with a delay of almost 8 days, but after a few days it stabilized at almost zero. Finally, YouTube had a minimal delay at the start and end, but in the middle there many fluctuations in the delay, reaching up to 4 days.}
    \caption{Platform-wise daily mean of the delay (measured in days) in the communication of statements of reasons to the \texttt{DSA-TDB}.}
    \label{fig:sor_communication_delay_timeseries}
\end{figure}

Regarding communication delay, LinkedIn, Pinterest, and Snapchat communicated their SoRs on the same day of application with little to no fluctuations, as shown in Figure~\ref{fig:sor_communication_delay_timeseries}. Despite the late start, X caught up on the first week of submissions by communicating content moderated since 25 September. Afterwards, it began to communicate SoRs the same day of their application. On the contrary, TikTok started by stably submitting on the same day of application, but in the second half of the 100-day period it communicated SoRs with an uprising mean delay of up to eight days. Facebook and Instagram communicated their SoRs with a mean delay of one day, with a few delay spikes common to both Meta platforms. On the other hand, YouTube had a minimum delay of one day, but with marked fluctuations throughout the period of analysis.

\begin{figure}[t]
    \centering
    \includegraphics[width=1\textwidth]{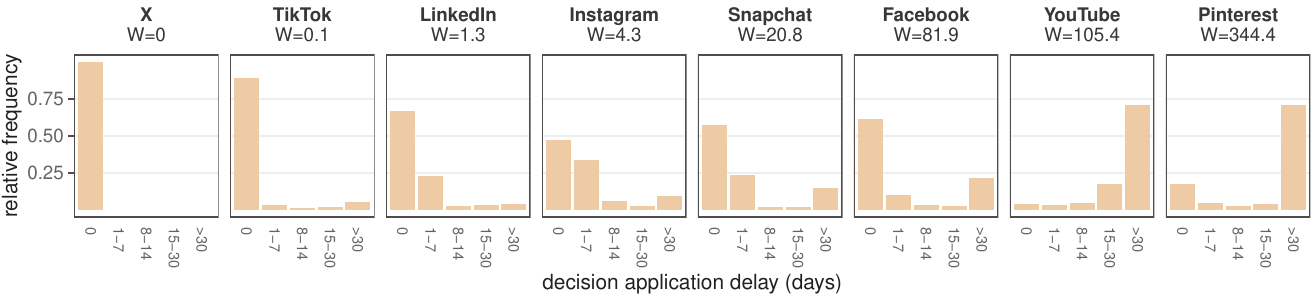}
    \Description{Faceted barcharts of the delay distribution. X is the only platform with a zero-day delay in all of its statements or reasons, with TikTok following closely. LinkedIn, Instagram, Snapchat, and Facebook have a zero-day delay in more than half of their application decisions. On the other hand, for both YouTube and Pinterest most of their delays are of 30 days or more.}
    \caption{Distribution of decision application delays (with respect to the date of content creation) by social media platform, ordered by their 80\% winsorized means (W).}
    \label{fig:sor_application_delay_trellis}
\end{figure}

Concerning application delay, we observed great heterogeneity in its distribution among platforms. For this reason, in their analysis and visualization we opted for using a few categorical cuts, as shown in Figure~\ref{fig:sor_application_delay_trellis}. In addition to visualizing the distributions of applications delay, we also report in the figure the 80\% winsorized mean (W) as a robust measure of centrality that lessens the influence of outliers. Oddly, X was the only platform that \textit{always} applied a decision on the same day the infringing content was created (W$=$0). TikTok had the second lowest delay, with 89\% of decisions being applied the day of content creation. Next, LinkedIn applied its decision within 7 days in 89.8\% of their SoRs, while Instagram in 81.7\%. Snapchat had a similar value for the same timeframe (81.4\%), but in 14.1\% of SoRs it took more than 30 days to apply its decision. Facebook presented a similar but more marked situation, with 72\% and 21.5\%, respectively. YouTube, on the other hand, had a substantial amount of decisions that took more than 30 days (70.7\%), while also having many within 15--30 days (17.5\%). However, the distribution with the highest mean and dispersion was that of Pinterest, as 17.5\% of decisions were applied the same day of content creation while 70.8\% after 30 days, with many even taking hundreds of days.

\subsection{Automated Means}
\label{sec:results_automated_means}

\begin{figure}[t]
    \centering
    \includegraphics[width=.925\textwidth]{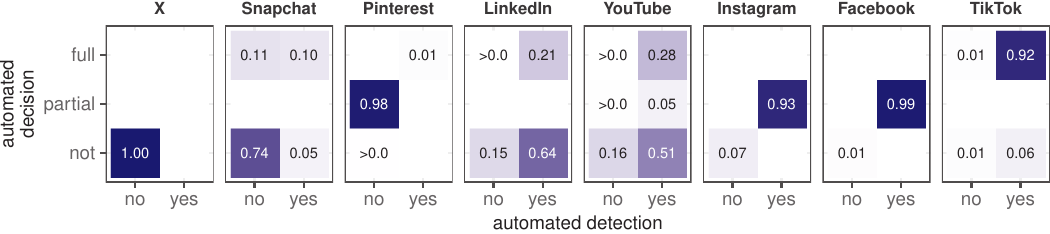}
    \Description{Faceted heatmap by platform, with rows indicating three automation decision levels (full, partial, and not) and columns indication automated detection (no or yes). At one extreme, X is the only platform that has all of its statements of reason as not using any automated means. At the other extreme, TikTok indicates that 92\% of its statements of reasons use full automated decision and detection means.}
    \caption{Platform-wise distribution of automated infringement detection and decision, ordered by an ad hoc automation index.}
    \label{fig:sor_automation_trellis}
\end{figure}

Automation can be applied at various steps in the content moderation process. It is often used to scale up the \textit{detection} of offending content and to speed up the \textit{decision} on how to moderate it. Consequently, the \texttt{DSA-TDB} includes two attributes to describe the possible use of automation for moderation (\S 9 Automated detection and \S 10 Automated decision): \textit{automated detection} (yes, no) and \textit{automated decision} (not, partially, fully). Figure~\ref{fig:sor_automation_trellis} provides a combined view of the use of automation for detection and decision by social media platforms. Again, we see different approaches being adopted. Surprisingly, \textit{none} of the SoRs submitted by X specified the use of automated means whatsoever. Snapchat is second in terms of non-automated means in their SoRs (74\%). Pinterest mainly uses non-automated detection with partially automated decision (98\%), whereas LinkedIn mainly uses automated detection together with non-automated decision (64\%). YouTube is the only platform that uses all combinations of automated means, with automated detection and non-automated decision being the most common (51\%), followed by fully automated means (28\%), and non-automated means (16\%). Once more, Instagram and Facebook have similar distributions, with automated detection and partially automated decision being the most common for both, with 93\% and 99\%, respectively. Almost opposite of X, TikTok predominantly indicates fully automated means for moderation (92\%).

\begin{figure}[ht]
    \centering
    \includegraphics[width=0.5\textwidth]{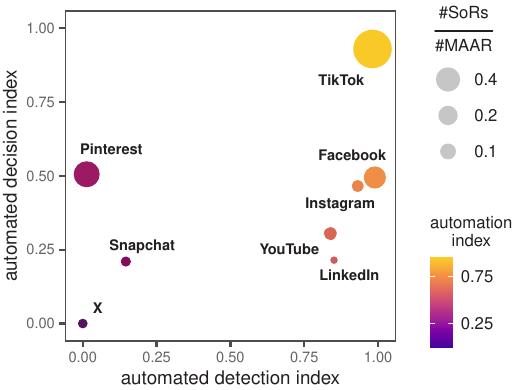} \Description{Scatterplot in which the eight platforms are distributed on a square panel based on their level of automated decision index and automated detection index. X, Snapchat, and Pinterest are on the quadrant for low automated means. LinkedIn, YouTube, Instagram and Facebook are on the quadrant with high detection but low automated detection. TikTok is the only platform present in the quadrant with high automation in terms of detection and decision. Incidentally, X and TikTok are almost diametrically opposite in terms of automated means.}
    \caption{Use of automated means by social media platforms for the detection of infringements and decision on these. Size is proportional to the number of statements of reasons (SoRs) per monthly authenticated active recipients (MAAR). In general, the bigger this ratio is, the higher the automation index ($r = 0.6)$, with Pinterest being somewhat of an outlier due to its marginal automated detection.}
    \label{fig:sor_automation_scatter}
\end{figure}

To provide a holistic view of social media behavior with respect to automated moderation, we converted the boolean and ordinal attributes of the \texttt{DSA-TDB} into two numeric indices of automated detection and automated decision. Both indices are defined in the [0, 1] range and are valued, for each platform, depending on the frequency with which automated detection and decision occurred in the platform's SoRs. We were then able to position each platform in the bi-dimensional space depicted in Figure~\ref{fig:sor_automation_scatter}, where dot size is proportional to the number of SoRs per MAAR, and dot color represents the mean of the automated detection and automated decision indices. The platform positioning within the newly defined automation space reveals four groups of platforms with overall similar degrees of automation: low  (X and Snapchat), modest (Pinterest), substantial (LinkedIn, YouTube, Instagram, and Facebook), and high (TikTok). At the platform level, the ratio in SoRs per MAAR is moderately correlated with the automation index ($r = 0.6)$. However, the correlation between number of SoRs per MAAR and automated detection is much lower ($r=0.3$) compared to automated decision ($r=0.9$). In particular, Pinterest stands apart from others platforms due to its very low level of automated detection but very high level of partially automated decision.

\subsection{Comparison with Transparency Reports}
\label{sec:results_transparency_reports}

Within the DSA framework there are multiple mechanisms to ensure transparency in content moderation, including the \texttt{DSA-TDB} that we analyzed so far, and the DSA Transparency Reports. Both resources cover similar aspects of the content moderation process, albeit at different granularities. As such, there is overlapping between the information contained in the two resources, which opens up the possibility to carry out thorough platform-wise comparisons. In this section we leverage this opportunity to carry out \textit{the first systematic comparison} between the data submitted by each social media to the \texttt{DSA-TDB} and that declared in the corresponding transparency report. In detail, for each platform and for each dimension of moderation (e.g., timeliness, automation, etc.), we look for the corresponding information in that platform's most recent DSA Transparency Report. We then compare the two figures to corroborate or challenge the self-declared information provided by each platform. This comparison serves as a validation step of the \texttt{DSA-TDB} data, and allows for spotting inconsistencies and discrepancies. 
However, in general, we were not able to compare each dimension of moderation for each of the eight platforms, due to differences in the data they provided in their transparency reports. Still, all platforms ---with the exception of Snapchat--- provided detailed information about the use of automated means for moderation, for which we found many inconsistencies. Further, in the case of X, we found inconsistencies across multiple aspects. In the following, we present the results of our comparisons, starting from the platforms for which we found less issues.

\textbf{Pinterest, Facebook, Instagram, LinkedIn.} Pinterest's self-reporting was overall congruous and detailed. Interestingly, it was the only social media to use three automation levels in their transparency report ---manual, hybrid, and automated--- similarly to the organization of the \texttt{DSA-TDB}. Unsurprisingly, Facebook and Instagram share similar transparency reports, based on Meta guidelines. In their reports, both state the heavy use of automated means, including full automation for both detection and decision. However, \textit{none} of their SoRs specified fully automated decisions, let alone full automation overall, as shown in Figure~\ref{fig:sor_automation_trellis}. For the reporting period 25 April -- 30 September 2023 (159 days), Meta platforms automatically removed the following organic content (i.e., non-paid): for Facebook, 46.68M pieces of content (93.94\% of the total), and for Instagram 75.11M (98.44\% of the total). These proportions of automatically removed content are consistent with their SoR submissions only if there is no distinction between fully and partially automated decision. LinkedIn's transparency report covers moderation actions occurred between 25 August and 30 September. It distinguishes between job posts and other content, with the former having a more stringent moderation policy. In its report, LinkedIn claimed to have removed all kinds of violating content detected automatically 99\% of the time, whereas for submitted SoRs it was 85\% ---a relatively small discrepancy.

\textbf{TikTok, YouTube, Snapchat.} TikTok's transparency report lacks several details and their data only covers the month of September 2023, shortcomings that are acknowledged therein. Based on the more detailed actions for removed content, we found a considerable automation discrepancy. The report specifies that 45\% of non-ad content was removed automatically, whereas in the \texttt{DSA-TDB} it was 95\%, more than double for a partially overlapping period. In the case of YouTube, even if it is the only platform to use all combinations of automated means without one being predominant ---in accordance with its transparency report--- we found intriguing anomalies. In the \texttt{DSA-TDB} we saw a remarkable spike in SoRs applied on 5 October 2023, reaching 732.4K. For reference, YouTube's daily median is 112.8K. Almost all of these are ads disabled because deemed outside the scope of the platform, but via non-automated means, which seems implausible. Furthermore, there were three similar spikes in mid December with even higher single-day volumes (>1.25M each) corresponding to a majority of SoRs having automated detection but non-automated decision. Unfortunately, with the data available we could not delve deeper into these anomalies. Snapchat's transparency report is somewhat peculiar, as they already had in place a namesake global transparency report, released twice a year. Hence, they only integrated this report via a document with some high-level information specific to the EU, in order to comply with the first mandatory DSA reporting period. Nevertheless, we noticed an odd inconsistency. DSA transparency reports must include data about \emph{Trusted Flaggers}: legal entities responsible for notifying platforms of illegal content~\cite{eu2020DSA}. As per Article 19(2) of the DSA, Trusted Flaggers are officially appointed by certain EU bodies. However, at the time the SoRs were submitted to the \texttt{DSA-TDB}, no Trusted Flagger had yet been designated. For this reason, no platform reported in their SoRs to have received notifications from Trusted Flaggers (\S 7.1 Information source), with the exception of Snapchat, for which we found a few hundreds SoRs with notification \emph{source type} set as Trusted Flagger. Upon further inspection we realized that Snapchat has a similar namesake internal program,\footnote{\url{https://values.snap.com/safety/safety-center}} unrelated to the DSA's one. This homonymy probably caused some confusion in Snapchat's reporting to the \texttt{DSA-TDB}, where they referred to their internal program instead of the intended one.
Since then, some Trusted Flaggers have been appointed pursuant Article 19(2) of the DSA.\footnote{\url{https://digital-strategy.ec.europa.eu/en/policies/trusted-flaggers-under-dsa}} Future analyses of the notifications submitted by Trusted Flaggers under the DSA could therefore be meaningful, but should be careful to discard Snapchat's initial unreliable data.

\textbf{X.} Finally, the transparency report by X includes very detailed data regarding its moderation actions for the period 28 August -- 20 October 2023 (54 days), with 56K actions being taken on content (36.5\% via automated means) and 2M actions on accounts (74.3\% via automated means). Yet, all of their SoRs submitted to the \texttt{DSA-TDB} are marked as not using any automated means. In addition, during the 100-day period of our study ---which partially overlaps with the transparency report's period--- X only submitted 466.4K SoRs, compared to 2.05M moderation actions declared in the transparency report. Those actions covered 14 violation categories, which are similar but not the same as the \texttt{DSA-TDB} categories. For content, \emph{violent speech} and \emph{abuse \& harrassment} were the most common: 25.5K (47\%) and 12.1K (22\%), respectively. For accounts, \emph{platform manipulation \& spam} and \emph{child sexual exploitation}: 1.89M (95\%) and 60K (3\%) violations, respectively. However, 99.8\% of all SoRs submitted by X to the \texttt{DSA-TDB} targeted \emph{synthetic media}, mostly \emph{pornography or sexualized content}, as shown in Figures~\ref{fig:sor_content_type_barchart} and~\ref{fig:sor_categories_trellis}, which appears to be inconsistent with X's transparency report. Moreover, all of X's SoRs in the \texttt{DSA-TDB} reported a decision application delay of zero days, without any use of automation, which again, seems very implausible.
 \section{Discussion}
\label{sec:discussion}

We first discuss the adequacy of the data submitted to the \texttt{DSA-TDB} (RQ1), next we examine inconsistencies in the submitted data (RQ2), and then we describe interesting differences and similarities among platforms (RQ3). Based on our findings concerning these aspects, we make suggestions on how the current structure of the database could be improved. Finally, we describe the main limitations of our study, and ponder on the potential implications of the database and the DSA in general on future online moderation research.

\subsection{RQ1: Adequacy of Submitted Data}
\label{sec:discussion_rq1}

We noticed several inadequacies in the data submitted by multiple social media platforms. In several cases the attributes in the database were conceived to allow platforms a higher level of granularity in characterizing a SoR, compared to what is actually done. For instance, there is a severe lack of detail regarding the types of infringements. The optional \emph{category addition} is practically never used, and the rich set of available keywords in \emph{category specification} is severely underutilized. Further issues arise because a single moderation action can fall under multiple categories. For example, the removal of sexual content could be categorized under \textit{pornography or sexualized content} and further specified as either \textit{adult sexual material} or \textit{image-based sexual abuse} (\S 16 Category \& Specification in the official documentation). However, in some cases the same content might also be categorized as \textit{scope of platform service}, which includes a \textit{nudity} subcategory, or as \textit{protection of minors}, under \textit{child sexual abuse material}. Although the documentation encourages platforms to be as specific as possible and to include multiple categories and specifications, this almost never happens. As a result, a certain degree of ambiguity exists in how different platforms chose to categorize the moderated content. This leads to challenges in interpreting \texttt{DSA-TDB} records, as it is currently impossible to distinguish whether two moderation actions were categorized differently due to actual differences in the moderated content or due to inconsistencies in how platforms reported their actions. In particular, the category \textit{scope of platform service} has a higher level of uncertainty due to its multiple unrelated subcategories. From the analysis of Figure~\ref{fig:sor_categories_trellis}, we note however that all platforms either indicated \textit{pornography or sexualized content} or \textit{scope of platform service} as the most frequent type of infringement. The presence of the \textit{nudity} subcategory in \textit{scope of platform service} could possibly explain this result, indicating a general and platform-independent tendency to prevalently moderate adult content. However, due to the missing optional fields that further specify infringement categories, particularly those more wide-ranging such as \textit{scope of platform service}, it is not possible to draw a definitive conclusion in this regard.

Similarly, platforms have the possibility to apply more than one of the four types of restriction decisions to a SoR (\S 3 The type of restriction(s) imposed), but none of the social media did so. In particular, it was surprising to see that no \emph{monetary} restriction decision was specified whatsoever, especially by platforms with well-established content monetization schemes subject to strong moderation, such as YouTube~\cite{ma2022m}. Likewise, \emph{content type} allows multiple values (\S 2 Specifications of the content affected by the decision), but in the great majority of instances only one was set. This is particularly problematic with \emph{synthetic media}, as it would be important to also know the kind of media of the generated content ---whether text, audio, video, or a combination of them~\cite{alam2022survey}. Again, while the official documentation explicitly suggests including as many types as are applicable to the content, this was almost never done. These widespread deficiencies not only diminish the utility of the \texttt{DSA-TDB} by limiting the depth of available data, but also compromise its commitment to transparency.

In general, the initial implementation of the SoR communication systems by social media platforms was not as comprehensive as it could and should be. Content language is an illustrating example (\S 2.4 Language of the content). The language in which moderators work is one of the key aspects that VLOPs must cover in their transparency reports ---an expected requirement given the multilingual policy to which the EU strives~\cite{gazzola2016multilingual}. Still, none of the social media platforms set the content language in their SoRs. From a technical point of view, most if not all of these very large social media already have mechanisms to register or automatically detect the language of a piece of content or account (e.g., via the user interface settings). Moreover, this attribute is essential for the platform, not only for moderation via manual our automated means~\cite{lees2022new}, but also to sell targeted ads and generate revenue~\cite{beauvisage2023online}. We presume platforms avoid communicating this attribute ---and the resulting workload--- simply because it is optional.

That being said, we believe that some fault also lies in the shortcomings of the current database schema. We acknowledge the challenges entailed in the definition of a common and unifying schema, not only for several social media, but for all the kinds of online platforms covered by the DSA. In fact, in mid 2023 there was a public consultation\footnote{\url{https://digital-strategy.ec.europa.eu/en/library/digital-services-act-summary-report-public-consultation-dsa-transparency-database}} on the \texttt{DSA-TDB}, in which 48 stakeholders participated, including representatives of 6 VLOPs. The consultation was intended to refine an existing draft of the schema and SoR communication mechanisms. This also means that VLOPs only had a few months to implement these changes. Nonetheless, we noticed that certain attribute definitions could be improved, based on their actual use. For instance, many platforms apply their restriction decisions at the account level, not because of a specific type of multi-media content created by it, but due to the account itself. However, in the initial database schema it is not possible to specify an account as the target of a restriction (\S 2.1 Type of content affected). Hence, many platforms indicated the SoR \emph{content type} as \emph{other} and detailed it with a free text label related to accounts, as we show in Figure~\ref{fig:sor_content_type_barchart}.

Such inadequacy of the database schema leads to confusion as to what is the actual target of a moderation action. Indeed, since the \emph{other} content type is detailed in a free text field with all sorts of heterogeneous labels, including various kinds of accounts and groups, we were unable to accurately distinguish among these for further analysis. The issue also propagates to other related database fields, such as the creation date of the infringing target. As indicated in the documentation (\S 2.3 Date on which the content was created on the online platform), this might be related to ``the date that specific content was posted, or the date that a user account was registered'' ---an attribute of two inherently different entities conflated into a single field. This and other ambiguities can have important repercussions on downstream analyses.
For instance, in Figure~\ref{fig:sor_application_delay_trellis} we show the histograms of the platform-wise distribution of the delays between the application date of a moderation action and the creation date of the corresponding moderated target. With the current \texttt{DSA-TDB} schema it is not possible to confidently assess whether the relatively high values on the extreme histogram bins (zero-day delay and more than 30 days delay) in some platforms (e.g., Facebook, Instagram, Pinterest) is due to the the inherently different creation times among different targets (e.g., content, account, groups), or due to some other phenomena.
This practical example underlines the implications of a poorly specified database schema and calls for additional specification effort.
Appropriately enough, the EU has undertaken an open approach to further develop the \texttt{DSA-TDB} and states it is willing to receive and act upon feedback.

\subsection{RQ2: Data Inconsistencies}

We found several inconsistencies in the self-reporting of social media platforms, based on a comparison between their DSA Transparency Reports and their SoRs. According to our comparative analysis, Pinterest resulted to be the most consistent in this regard, while also offering detailed data on its moderation actions. YouTube was also consistent in its self-reporting data, albeit the prominent spikes of restricted ad content in early October and mid December 2023 cast doubts on the completeness of the submitted data. The peculiar inconsistency of Snapchat, which indicated hundreds of notifications of illegal content by Trusted Flaggers (\S 7.1 Information source) when none had been designated under the DSA at the time, seems to be an honest mistake due its internal namesake program. Nonetheless this peculiarity stands out as an example of the complexity that emerges when adapting already established internal processes to new and vast norms.

The majority of inconsistencies that we found were related to the use of automated means (\S 9 Automated detection and \S 10 Automated decision).
In some cases, such inconsistencies appear to be a matter of interpretation. For instance, Facebook states in its transparency report: ``[e]very day, we remove millions of violating pieces of content and accounts on Facebook. In most cases, this happens automatically, with technology to detect, restrict, and remove content and accounts [...]''. Later, they further specify: ``when confident enough that a post violates one of our Community Standards, the artificial intelligence will typically remove the content or demote it. We also use artificial intelligence to select the content for human review on the basis of severity, virality, and likelihood of a violation''. Instagram's report includes similar statements. In other words, both Meta platforms declared heavy use of fully automated means in their transparency reports, but not a single fully automated action is reported in their submissions to the \texttt{DSA-TDB}, as shown in Figure~\ref{fig:sor_automation_trellis}. Instead, they opted to indicate partially automated detection by an overwhelming majority. Perhaps this choice was based on Meta's declarations of ``us[ing] human reviewers to assess the accuracy [of automated means] against current content, behaviour, or accounts, rather than just historical ones''. Regardless of the reason, we beg to differ, as we believe that at least some SoRs should have been classified as fully automated, because ---as implied by Meta itself--- there is no human involvement during the moderation action itself.

In other cases, the inconsistencies are a mismatch between the reported share of automated actions between the platforms' transparency reports and the \texttt{DSA-TDB}. Some inconsistencies are relatively small, as in the case of LinkedIn whose reported use of full automation differ by circa 10 percentage points. Other times the inconsistencies are much larger, as for TikTok, whose differences are in the region of 50 percentage points. 

Instead, X emerges as a distinctive outlier and as the most prominent case of self-reported inconsistencies. In its transparency report, the platform states: ``X employs a combination of heuristics and machine learning algorithms to automatically detect content that violates the X Rules and policies [...]''. Later they reiterate: ``[c]ontent flagged by these machine learning models are either reviewed by human content reviewers before an action is taken or, in some cases, automatically actioned based on model output''. Yet, \textit{all} of the SoR submissions by X during the period of study specified non-automated detection and decision. Not even a single partially automated decision was reported, as shown in Figure~\ref{fig:sor_automation_trellis}, which represents a striking inconsistency. In addition to the lack of use of automated means, other discrepancies include a lower than expected number of moderation actions, the unlikely predominance of illegal content, as well as a single value for both content type and restriction decision. These discrepancies in number and characterization cast serious doubts on the reliability of the SoRs submitted by X. These concerns appear to be well-founded, considering that during the same period in which we conducted our analysis, the European Commission initiated formal proceedings to evaluate whether X may have violated the DSA in aspects including, among others, content moderation practices \cite{eu_proceedings_ag_x}. On the other hand, X's transparency report is arguably one of the most, if not the most, comprehensive of the eight platforms, with detailed tables and explanations regarding their moderation actions.

\begin{figure}[t]
    \centering
    \includegraphics[width=.425\textwidth]{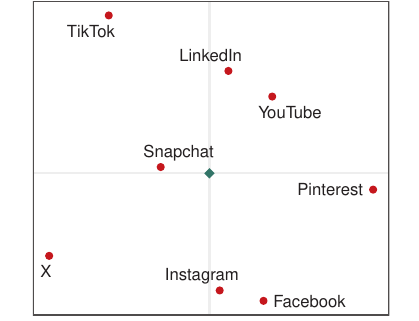}
    \Description{Square panel in which the centroid of the platforms' characteristics is almost at the center. X is the platform furthest from the centroid, The second furthest is TikTok, but in a different quadrant. Pinterest is the third most distant platform, in yet another quadrant. Facebook and Instagram are close to one another but somewhat distant from the centroid. LinkedIn and YouTube are closer to the centroid and also somewhat close to one another. Finally, Snapchat is the closest platform to the centroid. }
    \caption{Bi-dimensional Sammon mapping of very large social media platforms, based on the Manhattan distance of ten of their most relevant self-reported moderation metrics used in the study. The green diamond shape marks the platforms' centroid.}
    \label{fig:sor_sammon_mapping_scatter}
\end{figure}

\subsection{RQ3: Platform Differences}

As already mentioned, we found considerable differences in the way the platforms specify their SoR submissions to the \texttt{DSA-TDB}. For illustrative purposes, in Figure~\ref{fig:sor_sammon_mapping_scatter} we used ten metrics from the presented analyses to visualize the similarity among very large social platforms in terms of their MAAR and SoR submissions. We utilized Sammon mapping, a non-linear multidimensional scaling technique commonly used in exploratory data analysis~\cite{saeed2018survey}, setting the Manhattan distance as a similarity measure.

From this visualization, the dissimilarity of X with respect to the other platforms is evident. Indeed, we found that X greatly differed from the other platforms in terms of automated means, decision grounds, as well as content and restriction types and sub-types. However, we also found that X was the platform with more inconsistencies with respect to its DSA Transparency Report.
Pinterest is the second largest dissimilar platform, mainly due to their remarkable high average and dispersion in the delay between content creation and application of the moderation, as shown in Figure~\ref{fig:sor_application_delay_trellis}. Another difference lies in its high share of actioned \emph{pornography or sexualized content}. TikTok, somewhat diverges due its high volume in SoRs (52.69\% of all social media platforms), SoRs per MAAR, and predominant use of fully automated means, as reported in Table~\ref{tab:overview} and Figure~\ref{fig:sor_automation_trellis}. Whereas LinkedIn does so because it has the lowest volume of SoRs, MAAR, and SoR per MAAR. Unsurprisingly, Facebook and Instagram are the closest to each other. According to their transparency reports, they apply the same or very similar moderation policies, in line with Meta guidelines, which is reflected in their SoRs. Their main dissimilarity lies in the volume of moderation actions, with Facebook having almost ten times more SoRs for a similar number of MAAR than Instagram. Finally, Snapchat and YouTube share many characteristics with the other platforms, given the more balanced specification of their SoRs.

Differences among social media in terms of SoRs are not only the result of their moderation policies, but also, and perhaps foremost, due to the scope of the platforms themselves. Different social media cater for different users by means of their user experience, content offering, and community dynamics. Consequently, social media users usually browse more than one platform, each at distinct moments, for different durations, and with diverse motivations~\cite{alhabash2017tale}. These intrinsic differences should be taken into consideration when analyzing the content of the \texttt{DSA-TDB}. For instance, in our exploratory analysis we measured the seasonality strength of the number of daily interventions based on iterative STL (i.e., seasonal and trend decomposition using {\small LOESS}). LinkedIn had a very strong weekly seasonality of 0.73 on a 0 to 1 scale, almost double of the second one. In fact, we found a marked decline in LinkedIn SoRs with application date on weekends, while peaks usually occurred mid week. This is likely due to two factors: (\textit{i}) there is less human moderator activity on weekends ---which is relevant considering that only 21\% of LinkedIn decisions are automated--- and (\textit{ii}) users mostly use the platform on weekdays while at work ---given LinkedIn's focus on business and employment. Moreover, there is a marked decline in SoRs during the holiday season. This is an example of the peculiarities of the different platforms, and complementary data sources are needed to better understand this phenomenon. However, it serves the purpose of showing the great heterogeneity of platforms and of their moderation actions.

\subsection{Suggestions to Improve the DSA Transparency Database}

Despite the leap forward in terms of public accountability of online platforms brought by the \texttt{DSA-TDB}, its initial schema suffers some significant issues that hinder its usefulness for analysis and research.
As we have mentioned before, we are aware of the enormous challenge that is to develop a single database that covers the various moderation practices of many heterogeneous online platforms at a continental scale.
Based on our analyses, we have identified four issues that we believe should be fixed in future iterations of the database schema. To tackle these, we make the suggestions detailed in the following paragraphs, taking into account current practices of online platforms, the data inconsistencies that we found, the contents of the DSA itself, as well as the \texttt{DSA-TDB} documentation.

\subsubsection{Make more attributes for decision grounds mandatory}
As we previously detailed in Section~\ref{sec:discussion_rq1}, there are several attributes that can be used to describe the grounds for decision concerning a SoR, but for the most part these are underutilized, particularly the optional \emph{category addition} and \emph{category specification}.
Hence, we believe that the \texttt{DSA-TDB} should at least make mandatory the currently optional and multi-valued \emph{category specification}. This relatively small change to the database schema would oblige platforms to provide more context regarding their SoRs and it would greatly ease the analysis of grounds for decision, especially for values in \emph{category} that have wide-ranging descriptions, such as \emph{scope of platform service}.

\subsubsection{Differentiate among cause and target entities of moderation decisions}
In practice, moderation actions by online platforms usually involve three different entities: a single piece of content, an account, or a community group. We could say that a moderation decision is applied to a \emph{target} entity due to the infringement that an associated \emph{cause} entity represents.  In most instances, the cause and target entities of a decision are the same. Oftentimes, however, there is an overlap among these, such as when the removal of an account or community group is also accompanied by the removal of their contents. In the most common case, platforms apply a moderation decision to a single piece of content because of the piece itself. Other times, if the piece of content is deemed a severe violation, the decision is applied to the account in addition to the infringing content. Both of these cases are mostly covered by the current structure of the \texttt{DSA-DTB}.
Nonetheless, sometimes a decision is applied to an account by its behavior and not due to a specific piece of content, as in the case of repeated violations and the impersonation of people and companies not admissible under parody laws. In addition, a moderation decision might be applied to a community group without directly affecting the accounts that participated therein.
These other cases are not well covered by the current \texttt{DSA-TDB}, forcing platforms to awkwardly map them to the available attributes. For instance, as illustrated in Figure~\ref{fig:sor_content_type_barchart}, most of Snapchat's SoRs with \emph{content type} set to \emph{other} are accounts. Similarly, for Facebook and Instagram all \emph{other} content types are accounts, which represent more than half of their total SoRs.
Moreover, Article 17(1) of the DSA states that SoRs must be provided by platforms ``on the ground that the \emph{information} provided by the recipient of the service is illegal content or incompatible with their terms and conditions''. We argue that the term \emph{information} goes beyond the mere content created by the recipient ---e.g., it could be false information provided by the recipient when creating an impersonation account. Therefore, in spite of the relatively important impact on the database schema that such a differentiation among these entities represents, we believe that this change is justified, especially for what concerns differentiating between content and account.

\subsubsection{Flag synthetic content and accounts separately}
We argue that the current use of the value \emph{synthetic media} for the attribute \emph{content type} is semantically inappropriate and prone to overuse by online platforms, as was the case with X (see Figure~\ref{fig:sor_content_type_barchart}). In addition, accounts deemed to be synthetic might mainly interact with non-synthetic content, such as via content engagement botnets for hire, for which specifying synthetic content on the respective SoR might not be suitable. Instead, we think that the cause entity of a moderation decision (e.g., content, account) should have a separate boolean attribute \emph{synthetic}. Thanks to this relatively small change, it would be possible to better analyze the multimedia types that are used in contents and accounts deemed synthetic.

\subsubsection{Clarify and expand attributes for automated means}
We think that the two attributes related to the use of automated means (detection and decision) are not well specified, which gave way to different interpretations by the analyzed platforms, as we detailed in Sections~\ref{sec:results_automated_means} and~\ref{sec:results_transparency_reports}.
For instance, the official \texttt{DSA-TDB} auxiliary documentation titled \emph{Additional Explanation For Statement Attributes}\footnote{\url{https://transparency.dsa.ec.europa.eu/page/additional-explanation-for-statement-attributes}} reads in \S~9 Automated detection: ``This attribute automated\_detection indicates whether and to what extent automated means were used to identify the specific information addressed by the decision. `Yes' means that automated means were used to identify the specific information addressed by the decision''.
We argue that the current dichotomous attribute does not actually indicate to what extent automated means were used to identify infringing information. It merely indicates their use or not. Instead, we suggest using an ordinal attribute with the same values as \emph{automated decision}: fully automated, not automated, and partially automated. We also suggest a more detailed definition of the previous three values. In \S~10 Automated detection of the aforementioned auxiliary documentation, it is stated that ``‘Fully automated’ means that the entire decision-process was carried out without human intervention.'' We did not find, however, a definition of the \emph{decision-process} to which this statement refers within the \texttt{DSA-TDB} documentation, and we believe this to be the main source of variation in interpretation of automated means among platforms and even within the same platform. This seems to be the case of both Meta's platforms, which declare a heavy use of fully automated decisions in their transparency reports but afterward declared none to the \texttt{DSA-TDB}, opting instead to mostly submit SoRs as partially automated decisions.
We believe that the above decision-process should be delimited by interaction with the infringing entity of the SoR itself. If a human actor was not directly involved in the detection (e.g., by flagging) or decision (e.g., by reviewing) applied to the specific piece of content, account, or community group associated with the given SoR in which automated models or algorithms were used, then the respective means should be considered fully automated. This small change in schema and documentation would provide a consistent and more nuanced scale across both automated means. It would also reduce the current ambiguity regarding the related yet separate prior development (and refinement) process of the automated means themselves (e.g., algorithmic specification or training of AI models), which in most, if not all, cases involves a human actor at some point or another.

\subsection{Limitations}
Given the newness and complexities of the DSA in general and the \texttt{DSA-TDB} in particular, our study presents important limitations.
One is the lag of several months between the last available transparency reports we used and the 100-day period of study. Even in this relatively short time frame, it is possible that the user share of platforms or their moderation practices have changed noticeably, explaining in part some of the inconsistencies we reported. However, we believe that if that were the case, our overall findings would still hold. In addition, this limitation is intrinsic to the nature of the reports themselves, which are expected to be published a few months after the end of their reference period.
Another limitation related to the transparency reports is the lack of a standard format, which renders a direct comparison among them and with the data in the \texttt{DSA-TDB} particularly challenging, and even impossible in certain aspects due a lack of detail in some reports. In this regard, we think that DSA Transparency Reports should follow a similar structure, and at least use the common terminology already established for the \texttt{DSA-TDB}.
Another related important limitation concerns the lack of platforms' data outside the \texttt{DSA-TDB} to conduct data validation or more advanced analyses. For instance, and as already mentioned, we used MAAR as a proxy metric to relativize the volume SoRs, but it would have been more useful to also utilize the volume of pieces by content type created in the same time frame, which is not released by the eight analyzed platforms.
Furthermore, we initially intended to conduct more analyses on the SoR data, but as stated before, we realized upon exploration that the records were not filled in as expected, in addition to the several inconsistencies therein. Hence, we refrained from conducting additional analysis to avoid obtaining potentially misleading results. Among the fields that we did not consider, but that could be worth analyzing in future research, are the duration of the moderation actions (\S 4 The duration of the restriction), their territorial scope (\S 5 The territorial scope of the decision), and the free text description of the facts and circumstances that led to the moderation (\S 6 Description of the facts and circumstances).

\subsection{Implications for Future Research on Online Content Moderation}

The launch of the \texttt{DSA-TDB} marks a major milestone in regulation and accountability of online content moderation. Albeit it is still too early to confidently evaluate the impact of this database and the DSA as whole on industry, academia, and society in general, in the following paragraphs we ponder on some of the most important implications for research on online moderation, both in the medium and long terms.

\subsubsection{Unprecedented Access to Large Scale Moderation Data}
One of the key aspects of the \texttt{DSA-TDB} is the public availability of the full archive data, all for free and without restrictions. This means that anyone, including industry and scholarly researchers, could analyze these data to better understand the dynamics of online moderation in the EU. It should be noted, however, that the data on the \texttt{DSA-TDB} is intentionally not exhaustive, due to privacy concerns and the scope of the related Article 17 of the DSA, which only covers statements of reasons that platforms must provide to recipients when a restriction is applied to their information on the platform.
The DSA provides, however, a different mechanism for vetted researchers to gain access to more comprehensive data directly from VLOPs and VLOSEs by means of Article 40(12), subject to an application and approval process. As part of this vetting process, interested researchers must meet several requirements concerning, among other, their affiliation, research scope on systemic risk (as defined in Article 34) and/or risk mitigation (as defined in Article 35), ability to fulfil data security and confidentiality, and commitment to publicly release the derived research outcomes free of charge.
Notwithstanding the stringent requirements laid out in Article 40, such mechanism allows a more fine-grained and comprehensive scrutiny of the most prominent online platforms, and it could also be used to complement the data available on the \texttt{TSA-DBA}.
At the moment of writing (mid 2024), the affected social media platforms have already put in place their vetting process. Nonetheless, several researchers have raised concerns regarding the full compliance of the implemented processes~\cite{jaursch2024dsa}. Furthermore, the European Commission has initiated formal proceedings against the platform X, because ---among other preliminary findings--- it has so far failed to provide access to approved researchers as required by Article 40(12).\footnote{\url{https://ec.europa.eu/commission/presscorner/detail/en/IP_24_3761}}
The actual consequences for very large platforms that do not comply with the DSA and their effectiveness remain to be seen, however.

\subsubsection{Common Conceptual Model for Online Moderation}
It could be argued that the schema and documentation associated with the \texttt{DSA-TDB} codify a common core conceptual model to discuss online moderation across different kinds of platforms. Due to the large scope of the database, the definitions of the attributes and values therein are likely to be echoed across many derived scholarly works and beyond. Therefore, it is paramount that their description is clear and as unambiguous as possible for all the stakeholders involved in the production and consumption of these data. As we have shown in our analyses herein, this is not yet the case.

\subsubsection{Risk of Over-representing European Social Media Practices}
Several scholars have decried that despite their global reach, there is a prevalent US-centrism (and lesser Euro-centrism) in online social media platforms ---in policies, content created, and studies performed~\cite{nieborg2022platforms}.
In fact, all of the eight platforms analyzed herein were founded and are headquartered in the United States. In view of the previous two implications regarding data availability and a common conceptual model via the \texttt{DSA-TDB}, it is very likely that in the following years there will be a remarkable increase in the share of online moderation studies based on data and policies from within the EU.
Nonetheless, researchers and experts should refrain from generalizing the findings of these studies to other regions of the world, particularly low-income countries, which are usually under-represented in social media research but often consist of sizeable or even larger user bases, compared to most EU countries.

\subsubsection{A Word of Caution on Initial Database Analyses}
Based on our results, we conclude that great care is needed when conducting research studies that draw upon the initial data of the \texttt{DSA-TDB}. Above all, one must not forget that these data are self-reported. Even if SoR submission and transparency reporting are mandatory under the DSA, so far the reliability of the data lies on the willingness and the resources of the platforms to be comprehensive and accurate in their mapping of existing moderation processes to the novel and complex norms. As we have shown, this is not yet the case for the nascent \texttt{DSA-TDB}. Indeed, our analyses indicate that social media SoRs are only partially adequate with respect to the objectives of the database, manifest critical inconsistencies, and reflect platforms with very different scopes, domains, and dynamics. Moreover, data inconsistency and lack of data in optional attributes makes it difficult to perform more robust quantitative analyses to derive solid conclusions or generalize findings. Consequently, there is a high risk of making misleading inferences based on these data if the appropriate precautions are not taken, such as using complementary data sources for validation. Furthermore, it is possible that the issues described herein persist or are even exacerbated now that all providers of online services subject to the DSA ---and not only VLOPs and VLOSEs--- are obliged to submit their moderation data to the \texttt{DSA-TDB}~\cite{eu2020DSA}. Considering that large, resourceful, and supposedly well-organized platforms experienced the many issues discussed herein, we expect that smaller and less resourceful platforms will struggle to do any better. The impact of Article 17 of the DSA on smaller platforms is however yet to be determined.
 \section{Conclusions}
\label{sec:conclusions}

The recently established DSA Transparency Database is a trove of data on the moderation actions of large social media platforms, allowing the unprecedented opportunity to audit and compare their moderation practices. Here, we analyzed all 353.12M statements of reasons (SoRs) submitted by eight social media platforms during the first 100 days of the database. Other than providing one of the first accounts of the content of the database, our analyses also compared the reported moderation actions of the different platforms with what they declared in their transparency reports. We uncovered striking differences, inadequacies, and inconsistencies in the data submitted. Based on our results, we conclude that \textit{large social media adhered only in part to the philosophy and structure of the database}. Specifically, while all platforms met the requirements of the DSA, most omitted important (yet optional) details from their SoRs. This limits the usefulness of the database and hinders further analysis of this data by scholars and policymakers alike. At the same time, \textit{the structure of the database turned out to be partially inadequate} of the platforms' reporting needs. For example, the current structure lacks the possibility to explicitly report moderation decisions targeted at accounts rather than content. To this regard, our results can inform future developments of the database.

\textit{Social media platforms exhibited marked differences in their moderation actions}, in terms of restriction and content types, timeliness, and automation. Part of the differences are due to the platforms' varying degrees of adherence to the database. However, part are also due to the unique characteristics of each platform, which pose challenges when it comes to harmonizing their moderation practices, as envisioned by the DSA. Furthermore, we also found that \textit{a significant fraction of the initial database data is inconsistent}. This resulted from both internal validity checks on the data itself, as well as from external checks against statements and data extracted from the platforms' transparency reports. Of all the considered social media, \textit{X presents the most inconsistencies}, as reflected by striking discrepancies found in multiple aspects of its moderation actions. Finally, we conclude that the self-reported nature of the database, and the widespread inconsistencies that we found, \textit{raise concerns on the reliability of the data contained therein}. This begs caution when analyzing such data to make inferences and draw conclusions on the moderation practices of large social media platforms.

\begin{acks}

This work is partially supported by the European Union---NextGenerationEU within the ERC project DEDUCE (Data-driven and User-centered Content Moderation) under grant \#101113826, and the PRIN 2022 project PIANO (Personalized Interventions Against Online Toxicity) under CUP~B53D23013290006.
\end{acks}

\bibliographystyle{ACM-Reference-Format}
\bibliography{references}

\appendix

\section{Sources of DSA Transparency Reports}
\label{appx:dsa_report_links}
Table~\ref{tab:dsa_report_links} contains the URLs to the documents or repositories of the DSA Transparency Reports for the analyzed social media platforms. Beware that platforms differ in the periods of reference and publication dates in their reports. We used those that mainly focus on H1~2023 and were published around October of the same year. In some cases the active recipients of a given platform are detailed in a different document within the respective repository.

\begin{table}[H]
\caption{Sources of the DSA Transparency Reports of the analyzed social media platforms.}
\label{tab:dsa_report_links}
\small
\begin{tabular}{ll}
\toprule
\textbf{platform} & \textbf{URL of document or repository} \\ 
\midrule\addlinespace[2.5pt]
Facebook & \url{https://transparency.fb.com/reports/regulatory-transparency-reports} \\ 
Instagram & \url{https://transparency.fb.com/reports/regulatory-transparency-reports} \\ 
LinkedIn & \url{https://www.linkedin.com/help/linkedin/answer/a1678508} \\ 
Pinterest & \url{https://policy.pinterest.com/en/digital-services-act-transparency-report} \\ 
Snapchat & \url{https://values.snap.com/privacy/transparency/european-union} \\ 
TikTok & \url{https://www.tiktok.com/transparency/en/dsa-transparency} \\ 
X & \url{https://transparency.twitter.com/dsa-transparency-report.html} \\ 
YouTube & \url{https://transparencyreport.google.com/report-downloads} \\ 
\bottomrule
\end{tabular}
 \end{table}

\section{Daily Count of Moderation Actions}
\begin{figure}[H]
\centering
\includegraphics[width=1\textwidth]{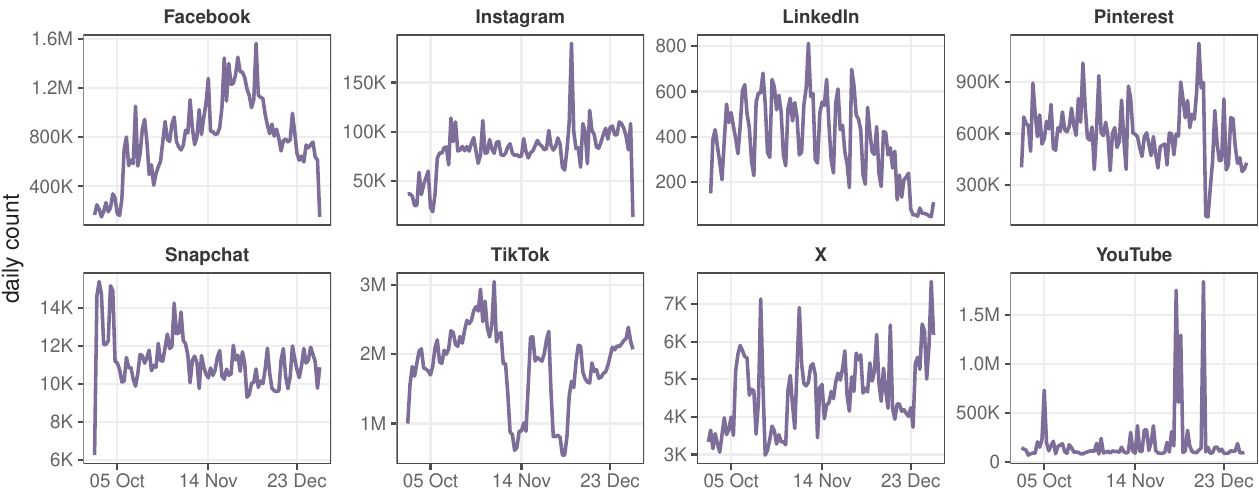}
\Description{Faceted time series. Facebook and Instagram reached their peaks round early December. Linkedin displays a weekly seasonality with a declining trend around the end of year holidays. Pinterest and Snapchat display a relatively stable trend over time. TikTok display two significant dips at the beginning of November and December, respectively.  YouTube presents three remarkable spikes. The lowest at the beginning of October, and then the two most prominent in early December.}
\caption{Platform-wise daily count of SoRs by application date of the restriction decision.}
\label{fig:sor_daily_application_date_timeseries}
\end{figure}

Figure~\ref{fig:sor_daily_application_date_timeseries} is the time series of the daily count of SoRs based on the application date of the decision for the action. Not too much attention should be paid to the values at the start and end of the time series. For instance, some platforms began their SoR communication a few hours late, which in part explains the seemingly low count at the start. In the case of Facebook and Instagram, they usually communicate their SoRs the day after their application date, hence the apparent sharp decline at the end of the period. The differences in daily volume of SoRs among the social media platforms is noteworthy. In this regard, there is a noticeable upward trend in Facebook, which began with a surge in moderation actions on 7 October, the day of the Hamas attack on Israel. A similar surge can be appreciated on Instagram, but with a relatively stable yet higher trend afterward. Despite their relevance, we did not delve into these phenomena because these are outside the scope of our study. Also of interest are: the strong weekly seasonality of LinkedIn and to a lesser degree that of Pinterest; the falls and rises on TikTok and X; and the YouTube spikes on early October and mid December, anomalies in terms of volume and characterization of SoRs for those days.

\received{January 2024}
\received[revised]{July 2024}
\received[accepted]{October 2024}

\end{document}